\documentclass[aps,preprint,floatfix,showpacs]{revtex4}
\usepackage{amsmath,amssymb,graphicx}

\begin{document}
\title{Stability of solutions of the Sherrington-Kirkpatrick model with
respect to replications of the phase space}

 \author{V.  Jani\v{s}}

\affiliation{Institute of Physics, Academy of Sciences of the Czech
Republic, Na Slovance 2, CZ-18221 Praha, Czech Republic }
\email{janis@fzu.cz}

\date{\today}

\begin{abstract}
We use real replicas within the Thouless, Anderson and Palmer construction
to investigate stability of solutions with respect to uniform scalings in
the phase space of the Sherrington-Kirkpatrick model. We show that the
demand of homogeneity of thermodynamic potentials leads in a natural way
to a thermodynamically dependent ultrametric hierarchy of order
parameters. The derived hierarchical mean-field equations appear
equivalent to the discrete Parisi RSB scheme. The number of hierarchical
levels in the construction is fixed by the global thermodynamic
homogeneity expressed as generalized de Almeida Thouless conditions. A
physical interpretation of a hierarchical structure of the order
parameters is gained.
\end{abstract}
\maketitle

\section{Introduction}
\label{sec:Intro}

The mean-field model for spin glasses introduced by Sherrington and
Kirkpatrick\cite{Sherrington75} is a paragon for complex statistical
systems. Although very simple in its formulation, the model offers almost
inconceivable richness of the phase space of its solution(s). This
richness is manifested in the replica-symmetry breaking (RSB) solution
introduced by G. Parisi within the replica trick.\cite{Parisi80} Since
then, a lot of supporting arguments reaching from numerical
simulations\cite{Binder86} over analytic thermodynamic approaches without
the replica trick\cite{Mezard87} to rigorous mathematical
constructions\cite{Guerra02} have been accumulated in favor of accuracy
and exactness of the Parisi RSB solution of the Sherrington-Kirkpatrick
(SK) model. In spite of the amassed evidence indicating to the RSB
character of the eventual solution of the SK model, we have not yet fully
understood in physical terms the origin of the RSB ansatz with its
ultrametric hierarchical structure of order parameters.

The replica-symmetry-breaking solution of Parisi was proposed as a means
for \textit{maximization} of the averaged free energy as a functional of
the averaged order parameters in the limit of zero number of mathematical
replicas. Analytic continuation from integer to non-integer numbers of
replicas less than one is, however, not trivial and unique. The maximum
principle seems to provide a way how to single out a particular analytic
continuation. Although the Parisi ansatz provides an internally consistent
solution numerically reproducing the results from Monte-Carlo
simulations,\cite{Kirkpatrick78} it is not evident whether it leads to the
absolute maximum of the free energy. Moreover, even when observed
empirically on simpler solutions of the SK model that more stable
solutions have higher averaged free energy, there is no general physical
law from which we could derive the maximum principle for the averaged free
energy. On the other hand, a supremum from all possible choices of the
Parisi order parameters $q(x)$, $x\in[0,1]$ was proved to lead to an exact
averaged free energy of the SK model.\cite{Talagrand04} The RSB scheme
hence has a deeper meaning and there must be a fundamental physical
principle from which one could derive the RSB solution without additional
physically unjustified ansatzes.
 
The aim of this paper is to demonstrate that the maximum principle in the
Parisi solution can be replaced by minimization of inhomogeneity of
thermodynamic potentials in a successive way toward a globally
thermodynamically homogeneous solution. Thermodynamic homogeneity is a
fundamental property needed for the existence of a unique thermodynamic
limit of statistical systems. It is a consequence of scale invariance of
the limit of volume $V$ of the system to infinity. That is, large
volumes $V$ and $\alpha V$ of thermodynamically homogeneous systems must
produce the same thermodynamics, i. e., the same densities of extensive
thermodynamic variables. Only in thermodynamically homogeneous systems the
thermodynamic limit does not depend on the shape and the boundary
conditions of large finite volumes.
 
We find it useful to apply specific scalings of extensive variables of
mean-field, long-range models represented by replications of the phase
space. We employ real replicas of the spin variables and demand that the
thermodynamics in the replicated phase space be independent of the number
of introduced equivalent replicas. The independence of the resulting
averaged free energy density on the number of real replicas is
investigated by studying stability of thermodynamic potentials with
respect to perturbations induced by infinitesimal homogeneous interactions
between different replicas. Thermodynamic potentials are stable if linear
response to the inter-particle interaction remains finite and the spin
replicas decouple in the equilibrium state after switching off the external
inter-replica interaction.

The role of real replicas in this approach is similar to the role of
mathematical replicas in the replica trick. They are used to represent
integer powers of the partition sum. Unlike the replica trick the number
of real replicas will not be limited to zero. Alike the replica trick we
will need to continue analytically the averaged replicated free energy
from integer numbers of real replicas to arbitrary positive numbers to
test thermodynamic homogeneity locally. For this purpose we will need to
assume a symmetry of the averaged replica-dependent order parameters,
Legendre conjugates to the inter-replica interaction. To find a physical
motivation for a selection of a particular symmetry, we use the
thermodynamic approach of Thouless, Anderson and Palmer (TAP). In this
approach we are endowed with a set of thermodynamic mean-field parameters,
local magnetizations $m_i$. They have to determine equilibrium
thermodynamic states at fixed configurations of spin-spin couplings. If
the full set $m_i, i= 1,\ldots, N$ determines the thermodynamic state
uniquely, the replica symmetric ansatz applies as in the high-temperature
phase. If not and the set of local magnetizations does not contain full
information about equilibrium states, the system is thermodynamically
inhomogeneous and further knowledge of the system is needed. We propose in
this paper a systematic way how to retrieve the missing information about
the structure of degenerate states described by a set of local
magnetizations. Our construction leads in a rather direct way to an
ultrametric structure of the order parameters in the (real) replica
indices and to a hierarchical averaged free energy equivalent to the
discrete RSB solution of Parisi.

\section{Thermodynamic homogeneity and averaging of replicated TAP
free energies} \label{sec:TAP-replicated}

\subsection{Thermodynamic homogeneity and replications of the phase
space} \label{sec:homogeneity}
 
Homogeneity of thermodynamic potentials is one of basic principles of
statistical mechanics. Thermodynamic homogeneity in systems with
short-range interactions is usually expressed as the Euler condition for
thermodynamic potentials (free energy) $\alpha F(T,V,N,\ldots,X_i,\ldots)
= F(T, \alpha V, \alpha N,\ldots,\alpha X_i,\ldots)$, where $\alpha$ is an
arbitrary positive number and $X_i$ exhaust all extensive variables. Only
if the Euler homogeneity is fulfilled we are able to factorize the volume
from extensive variables, come over to densities, and define the
thermodynamic limit uniquely and independently of the shape and boundary
conditions of finite volumes. Thermodynamic homogeneity can be rephrased
as a scale invariance of entropy $S(E) = k_B \ln \Gamma(E) = k_B/\nu \ln
\Gamma(E)^\nu$ for arbitrary positive $\nu$. This definition extends also
to mean-field (long-range) models. We hence use the latter form of
thermodynamic homogeneity applied to the averaged free energy of the SK
model.
 
Assuming thermodynamic homogeneity we can write the averaged free energy as
$F  = -1/\beta\nu \left\langle\ln\left[\text{Tr}\ e^{-\beta
H}\right]^\nu\right\rangle_{av}$. If the scaling factor $\nu$ is a
positive integer we can equivalently represent the discrete multiplication
of the phase space via replicating the dynamical variables in the
partition sum (folding of the phase space): $\left[\text{Tr}\ \exp\{-\beta
H\}\right]^\nu = \text{Tr}_\nu \exp\left\{\sum\limits_{a=1}^{\nu}
\sum_{<ij>} J_{ij}S_i^a S_j^a\right\}$. Each replicated spin variable
$S_i^a$ is treated independently, i.~e., the trace operator
$\text{Tr}_\nu$ operates on the $\nu$-times replicated phase space.
Calculation of the free energy in the expanded phase space amounts to
evaluation of the free energy of the replicated Hamiltonian. This
multiplication of the number of dynamical variables is called real
replicas and has been occasionally used, mostly to illustrate the meaning
of the overlap order parameters in the Parisi RSB
construction.\cite{Parisi83,Janis87,Franz93} Note that replicating the
phase variables $\nu$-times is not the same operation in long-range models
as a scaling of the volume $V\to\alpha V$. The spin-spin couplings, that
are in short-range models intensive variables, depend in long-range models
on the volume as well and are to be scaled, in the SK model as $J_{ij}\to
J_{ij}/\sqrt{\alpha}$, to compensate for additional couplings in the
inflated volume. Replicating of the phase space seems a more suitable and
simpler tool for investigating thermodynamic homogeneity of mean-field
models than direct scalings of the phase space with all new spins coupled
to the old ones. When we replicate the original phase space we completely
decouple the new replicated spins from the original ones and do not
thereby change the normalization of the spin-spin couplings. Moreover,
replication of phase variables is more suitable for investigating
stability with respect to perturbations induced by interactions between
different replicas without breaking translational invariance.
 
Real replicas are also of principal importance for the thermodynamic
construction of a mean-field theory of spin glasses, since they offer a
space for new symmetry-breaking fields. The real replicas are
independent when introduced. We break their independence by switching
on a (homogeneous) infinitesimal interaction between the replicas that we
denote $\mu^{ab}$. We then add a small interacting part $\Delta H(\mu)=
\sum_i \sum_{a < b}  \mu^{ab} S_i^a S_i^b$ to the replicated spin
Hamiltonian.  The averaged free energy per replica of the
 system with weakly interacting replicas reads
\begin{equation}\label{eq:avFE}
  F_\nu (\mu) = - k_BT\ \frac 1\nu
  \left\langle\ln\text{Tr}\exp\left\{-\beta\sum_\alpha H^\alpha -\beta
      \Delta H(\mu) \right\} \right\rangle_{av}\ . \end{equation}
The inter-replica interactions $\mu^{ab}>0$ play the role of
symmetry-breaking fields in the SK model. They induce new order parameters
in the response of the system to this field that need not vanish in the
low-temperature phase, when the linear response theory breaks down. They
allow to disclose the degeneracy when mean-field solutions do not
represent unique pure equilibrium states. The inter-replica interactions
are unphysical (not measurable) and hence to restore the physical
situation we have to switch off these fields at the end. If the system is
homogeneous we must end up with an identity
\begin{equation}\label{eq:av-homogeneity}
\frac{d}{d\nu}\lim_{\mu\to0}F_\nu(\mu) \equiv 0 \ .
\end{equation}
This quantification of thermodynamic homogeneity, thermodynamic
independence of the scaling parameter $\nu$, will lead us in the
construction of a stable solution of the SK model.

\subsection{Averaging of the replicated TAP free energy}
\label{sec:}

Thermodynamic homogeneity can be investigated in the SK model either in the
replica trick or in the thermodynamic TAP approach. Thermodynamic
homogeneity in the replica trick is equivalent to scale invariance of the
limit of the number of mathematical replicas to zero, i. e., the result
should be invariant with respect to scalings of the replica index
$n\to\alpha n$.\cite{Janis04a} We prefer to use here the thermodynamic TAP
approach so that to demonstrate that the RSB scheme is neither part of the
replica trick nor a consequence of the limit of the number of mathematical
replicas to zero. We also find the TAP approach more appropriate for
finding a physical interpretation of the role of the replicated spins.
They are used to lift degeneracy in the determination of equilibrium
thermodynamic states from the mean-field local magnetizations.

The TAP free energy can suitably be represented as
\begin{multline}\label{eq:TAP} F =\sum_i m_i\eta_i  - \frac 14\sum_{i,j}
\beta J_{ij}^2\left(1 - m_i^2\right)\left(1-m_j^2 \right) - \frac
12\sum_{i,j} J_{ij}m_i m_j \\ - \frac 1{\beta} \sum_i \ln
2\cosh\left[\beta\left(h + \eta_i\right)\right]\ .
\end{multline}
We introduced fluctuating internal magnetic fields $\eta_i$ as variational
parameters making free energy \eqref{eq:TAP} together with the local
magnetizations $m_i$ extremal. The stationarity equations for the internal
magnetic fields and for the local magnetizations read $\eta_i = \sum_j
J_{ij}m_j - m_i \sum_j \beta J_{ij}^2(1 - m_j^2)$, $m_i = \tanh[\beta(h +
\eta_i)]$, respectively. We can now try to solve these TAP equations for
$m_i, \eta_i$ on finite lattices with fixed configurations of spin-spin
couplings $J_{ij}$. We are then confronted with a plethora of metastable
solutions that are difficult to handle. Instead of individual solutions we
can better deal with the so-called complexity of the TAP equations, being
proportional to the total number of solutions.\cite{Bray80}

The existence of many metastable solutions of the TAP equations generally
hinders direct averaging over the random configurations of the spin-spin
couplings. If the direct method is used, that is if we remain within the
linear response theory with the fluctuation-dissipation theorem valid, we
end up with the SK solution.\cite{Sommers78,Janis89} M\'ezard et
al\cite{Mezard86} purposed the so-called cavity method to include ensembles
of statistically weighted TAP solutions into the averaging process and
succeeded in going beyond the SK solution toward the Parisi RSB scheme.
 
In this paper we want to avoid any special ansatzes about the structure or
the distribution of the TAP solutions and to remain entirely within the
direct averaging scheme with the ergodic and fluctuation-dissipation
theorems obeyed. The fluctuation-dissipation theorem, expressed in the TAP
construction as $\chi_{ii} = (1 - m_i^2)/T$, strictly holds only if the
individual TAP solutions determine unique thermodynamic states. It was
shown by Plefka that this is the case if $1 \ge \beta^2(1 - 2\langle
m_i^2\rangle_{av} + \langle m_i^4\rangle_{av})$.\cite{Plefka82} We,
however, know that this condition is violated in the low-temperature
phase, which led Plefka to modifications of the TAP
equations.\cite{Plefka02}

We in principle follow an analogous way to Plefka and assume that local
magnetizations determined from the TAP equations do not contain exhaustive
information about the equilibrium thermodynamic states. We modify the TAP
equations and connect violation of the Plefka condition with violation of
thermodynamic homogeneity. We introduce real replicas into the TAP
approach to substantiate this. We use the TAP free energy with $\nu$
equivalent spin replicas on each site. Real replicas were
introduced into the TAP approach from a different motivation by the author
years ago.\cite{Janis90} The result, a generalized TAP free energy with
$\nu$ replicas with switched off inter-replica interactions, $\mu^{ab}=0$,
can then be overtaken from Ref.~\cite{Janis90}. It reads
\begin{multline}\label{eq:nuTAP} F_\nu =\frac 1\nu
\sum_{a=1}^\nu\left\{\sum_i m_i^a\left[\eta_i^a + \beta J^2\sum_{b=1}^{a-1}
\chi^{ab}m_i^b\right] + \frac{\beta J^2N}2 \sum_{b=1}^{a-1}
(\chi^{ab})^2\right. \\ \left. - \frac 14\sum_{i,j} \beta J_{ij}^2\left[1
- (m_i^a)^2\right]\left[1-(m_j^a)^2 \right] - \frac 12\sum_{i,j}
J_{ij}m_i^a m_j^a \right\} \\ - \frac 1{\beta\nu} \sum_i \ln
\mbox{Tr}\exp\left\{ \beta^2 J^2\sum_{a < b}^\nu \chi^{ab}S_i^aS_i^b
+\beta\sum_{a=1}^\nu\left(h+\eta_i^a\right) S_i^a\right\}\ .
  \end{multline}
Here $m^a_i$ are local magnetizations and $\eta^a_i$ are local internal
magnetic fields. They are configurationally dependent variational
variables determined from stationarity equations. Parameters
$\chi^{ab}, a\neq b$, are overlap susceptibilities and are global
(translationally invariant) variational variables,
Legendre conjugates to the symmetry breaking fields $\mu^{ab}$. They
are the genuine order parameters in the spin glass phase of the SK
model in this construction.  At the saddle point we have $\chi^{ab}=
N^{-1}\sum_i\left[\langle S^a_iS^b_i\rangle_T - \langle
  S^a_i\rangle_T\langle S^b_i\rangle_T\right]$.

Free energy $F_\nu$ from Eq.~\eqref{eq:nuTAP} is averaged over thermal
fluctuations for one configuration of spin-spin couplings $J_{ij}$. To
perform averaging over the randomness in the spin-spin couplings we have
to decide whether the solutions of the replicated TAP equations,
stationarity equations derived from free energy \eqref{eq:nuTAP},
determine unique equilibrium thermodynamic states or not. If not, we have
to surmise the internal structure of equilibrium states represented by a
set of local magnetizations and averaged overlap susceptibilities. It can
be done only via an ansatz. The pure states in spin glasses are, however,
peculiar in that respect that they cannot be singled out by external
symmetry-breaking fields. To avoid application of any unjustified
ansatzes, we assume that the solutions of the replicated TAP equations do
represent unique thermodynamic states as it is the case in the
high-temperature phase. It means that replication of the phase space
serves as a replacement of symmetry breaking fields. Replicating the phase
space enable us to extend the high-temperature properties, that is the
replica symmetric ansatz, to low temperatures in analogy to the
ferromagnet in an external magnetic field. Then the linear response,
ergodic and fluctuation-dissipation theorems hold and we can use the same
averaging of the replicated TAP free energy as used to derive the SK
solution from the TAP free energy.

Even with the assumption of uniqueness of equilibrium states in the
replicated phase space the averaging of the replicated free energy
\eqref{eq:nuTAP} cannot be performed explicitly. We first have to quantify
thermodynamic equivalence of the replicated spin variables. Since the
replicated spin variables were introduced in the TAP approach to deal with
a possible degeneracy of solutions of the TAP equations properly, we assume
the following thermodynamic equivalence of real replicas motivated by  the
paramagnetic solution
\begin{equation}\label{eq:equivalence-replicas}
m_i^a \equiv \langle S_i^a\rangle_T = m_i \ .
\end{equation}
There is no apparent reason for breaking this equivalence in the spin glass
phase, since each copy of the spin variables shares the same external
macroscopic parameters determining the thermodynamic state.
Equation~\eqref{eq:equivalence-replicas} expresses the fact that each TAP
solution with local magnetizations $m_i$ represents on average $\nu$
thermodynamic states labeled by the replica index~$a$. The different
thermodynamic states are indistinguishable at the level of local
magnetizations. Since the replicated spin variables are distinct, their
local overlap susceptibility is not, however,  determined from the
fluctuation-dissipation theorem and enters the free energy as a
variational parameter. Equivalence of spin replicas,
Eq.~\eqref{eq:equivalence-replicas}, leads then to independence of local
magnetic fields $\eta_i^{a}$ and the sum of the averaged overlap
susceptibilities $\sum_b\chi^{ab}$ of the replica index $a$. With this
conclusion we can write down an explicit representation of the averaged
free energy density with $\nu$ equivalent real replicas
\begin{multline}\label{eq:FE-averaged-finite}
f_\nu = \frac{\beta J^2}{4} \left[\frac 1\nu\sum_{a\neq b}^\nu
\left\{\left(\chi^{ab}\right)^2 + 2 q\chi^{ab}\right\}
- (1 - q)^2\right]\\ -\frac
1{\beta\nu}\!\int\limits_{-\infty}^{\infty}\frac{d\eta}{\sqrt{2\pi}} \
e^{-\eta^2/2} \ln \text{Tr} \exp\left\{\beta^2J^2\sum_{a < b}^\nu
\chi^{ab}S^aS^b + \beta \bar{h}\sum_{a=1}^\nu S^a\right\}\ .
\end{multline}
We denoted the fluctuating magnetic field $\bar{h} = h + \eta\sqrt{q}$. The
averaged order parameters are at the saddle point $q = \langle\langle
S^a\rangle_T^2 \rangle_{av}$ and $\chi^{ab} = \langle\langle S^a
S^b\rangle_T\rangle_{av} - q$.

The trace in the averaged free energy density \eqref{eq:FE-averaged-finite}
cannot be evaluated explicitly. To do so we have to know the matrix of the
overlap susceptibilities $\chi^{ab}$ reflecting the structure of
thermodynamic states indistinguishable in the TAP equations.
Mathematically it means to find the most general structure of matrix
$\chi^{ab}$ with the constraint that $\sum_b \chi^{ab}$ does not depend on
$a$. Since we do not know how such a structure should look like, we have to
make a choice and check only a posteriori, whether our choice has led to a
consistent solution. We use an iterative construction with successive
replications of the phase space accompanied by the replica-symmetric
ansatz for the overlap susceptibilities at each step. We replicate the
system so many times until a thermodynamically homogeneous solution
satisfying Eq.~\eqref{eq:av-homogeneity} is reached.

\subsection{Analytic continuation and local and global thermodynamic
homogeneity} \label{sec:Local&global}

The mean-field (saddle-point) equations for $\chi^{ab}$ derived from the
averaged free energy density \eqref{eq:FE-averaged-finite} are identical
for all pairs of the replica indices $(ab),a\neq b$. There is no apparent
symmetry breaking force in the replica space and $\chi^{ab} = \chi$ is
evidently a possible solution. We choose this simplest, replica symmetric
solution so that we can evaluate the averaged free energy explicitly. This
replica-symmetric choice corresponds physically to a situation where the
TAP solutions comprise of equivalent distinguishable thermodynamic states
with the same overlap susceptibility (distance) between each pair of
different states. Hence no internal structure of equilibrium states is
assumed.

It is straightforward to evaluate the averaged free energy density $f_\nu$
with the ansatz $\chi^{ab} = \chi$. We employ the Hubbard-Stratonovich
transformation to decouple the replicated spins and end up with
\begin{multline}\label{eq:FE-1RSB}
f_\nu(q,\chi) = -\frac \beta4(1-q)^2  + \frac \beta4(\nu - 1)\chi(2 q +
\chi) + \frac\beta2 \chi- \\ -\frac 1{\beta\nu}
\int_{-\infty}^{\infty}\mathcal{D}\eta\ln\int_{-\infty}^{\infty}
\mathcal{D}\lambda  \left\{2\cosh\left[\beta\left(h +
\eta\sqrt{q} +\lambda\sqrt{\chi}\right)\right]\right\}^{\nu}
\end{multline}
where we used an abbreviation for the Gaussian differential
$\mathcal{D}\phi \equiv {\rm d}\phi\  e^{-\phi^2/2}/\sqrt{2\pi}$. We put
$J=1$ and use this energy scale throughout the rest of the paper.

The averaged free energy density, Eq.~\eqref{eq:FE-1RSB}, however,
apparently depends on $\nu$ whenever the order parameter $\chi >0$. It is
the case in the spin-glass phase below the de Almeida-Thouless (AT)
instability line. Free energy~\eqref{eq:FE-1RSB} has the form identical
with the Parisi one-step RSB solution, where $\nu$ plays the role of the
parameter dividing the replica space in the RSB ansatz.\cite{Parisi80} We
can easily analytically continue the r.h.s. of Eq.~\eqref{eq:FE-1RSB} to
all real numbers. The integral representation in
Eq.~\eqref{eq:FE-1RSB} is well defined and analytic for $\nu\in
(-\infty,\infty)$.

There are two observations we can make from the analysis of free
energy~\eqref{eq:FE-1RSB}. First, at any value of $\nu$ the overlap
susceptibility is positive below the AT line. It indicates that the SK
solution ($\chi=0$) becomes thermodynamically inhomogeneous. Second, we
are unable to find parameters $q,\chi$ for which free
energy~\eqref{eq:FE-1RSB} would be $\nu$-independent and hence the TAP
solutions indeed do not describe unique thermodynamic states in the
low-temperature phase. Our modification of the TAP free energy becomes
nontrivial.

Free energy~\eqref{eq:FE-1RSB} is not globally thermodynamically
homogeneous, since it depends on the scaling parameter $\nu$. We can,
however, optimize the solution in that we demand that the deviations from
the thermodynamic homogeneity be minimal. This is achieved if at
least thermodynamic homogeneity is obeyed locally, that is, if
\begin{equation}\label{eq:local-homogeneity}
\frac {\partial f_\nu(q,\chi)}{\partial \nu} =0.
\end{equation}
This equation determines an optimal parameter $\nu_{opt}$ for which the
free energy is locally thermodynamically homogeneous. We show later on
that Eq.~\eqref{eq:local-homogeneity} has always a solution with
$\nu_{opt} > 0$. Free energy~\eqref{eq:FE-1RSB} together with the
optimization condition~\eqref{eq:local-homogeneity} exactly deliver the
thermodynamics of the Parisi one-step RSB. Monasson and later M\'ezard
proposed in Refs.~\cite{Monasson95,Mezard99} a similar approach to the
thermodynamics of structural glasses, the so called cloned liquid. The
local homogeneity, Eq.~\eqref{eq:local-homogeneity}, was interpreted there
as vanishing of complexity.
 
Satisfying thermodynamic homogeneity locally for the optimal parameter
$\nu_{opt}$ is generally insufficient. We in fact should construct a
theory being globally thermodynamically homogeneous. To check whether free
energy~\eqref{eq:FE-1RSB} can for any $\nu$ be globally thermodynamically
homogeneous we have to perform a further scaling of extensive variables via
replicating the spin variables in Eq.~\eqref{eq:FE-averaged-finite}. We do
so by replacing $\nu\to n\nu$ and testing homogeneity of the free energy
$f_\nu$ with respect to the $n$-times enlarged (replicated) phase space.
With the new scaling we have to replicate each spin variable $S^a$ to
$(S^\alpha)^a$ and transform the matrix of overlap susceptibilities to a
super matrix $\chi^{ab}\to(\chi^{\alpha\beta})^{ab}$ where $a,b =
1,\ldots,\nu$ and $\alpha,\beta = 1,\ldots,n$. With this replication we
allow that the local spin variables $S^a$ may still be insufficient to
determine unique thermodynamic states. Since all spin variables $S^a$ are
thermodynamically equivalent, they have to split into new states labeled
by the new replica index $\alpha$ identically.

The mean-field equations for $(\chi^{\alpha\beta})^{ab}$ again contain a
replica-symmetric solution $\chi_1 = (\chi^{\alpha\beta})^{ab}$ for $a\neq
b$ and $\chi_2 = (\chi^{\alpha\beta})^{aa}$ for $\alpha\neq \beta$. This
symmetric solution assumes that the local TAP magnetization $m_i$ can be
represented by $\nu$ composite states each of which contains $n$ pure
states. Two pure states are distinguished by the overlap susceptibility
$\chi_2$ if they peel off from the same parental spin $S^a$ and by
$\chi_1$ if they stem from two different parental spins $S^a$, $S^b$,
$a\neq b$.

It follows from the presented construction that $\chi_1 \ge \chi_2$. The
free energy now generally depends on $n, \nu$ and $\chi_1,\chi_2$. We
obtain explicitly
\begin{multline}\label{eq:FE-2RSB}
f_{\nu,n}(q,\chi_1,\chi_2) = -\frac \beta4(1-q)^2 + \frac \beta 2 \chi_1 +
\frac \beta4\left[(\nu - 1)\chi_1(2q+\chi_1) + \nu (n -
1)\chi_2(2q+\chi_2)\right] \\ -\frac 1{\beta \nu n} \int\mathcal{D} \eta
\ln\int\mathcal{D}\lambda_1\left[\int\mathcal{D}\lambda_2\left\{
2\cosh\left[\beta\left(h + \eta\sqrt{q} +\lambda_1\sqrt{\chi_1-\chi_2} +
\lambda\sqrt{\chi_2}\right)\right]\right\}^{\nu}\right]^{n}\ ,
\end{multline}
which again is equivalent to the Parisi 2RSB free energy. Free energy
$f_{\nu,n}$ from Eq.~\eqref{eq:FE-2RSB} is independent of $n$ and reduces
to $f_\nu$ from Eq.~\eqref{eq:FE-1RSB} if and only if $\chi_2 =0$.

When this "second level" free energy $f_{\nu,n}$ depends on $n$ it can be
optimized so that it be locally thermodynamically homogeneous. The result
can be subject to a further scaling of extensive variables in order to
verify whether the resulting thermodynamic potential is globally
homogeneous. We can proceed with the hierarchical (multiplicative)
scalings accompanied with the replica-symmetric ansatz about the structure
of the newly generated overlap susceptibilities so long until the free
energy becomes a thermodynamically homogeneous function, i.~e., the free
energy does not depend on the last scaling parameter. It is evident that
with each transformation we generate just one scaling parameter $m_l$ and
one block-off-diagonal matrix of the overlap susceptibility $\chi_l$.

\section{Hierarchical mean-field theory}
\label{sec:Hierarchical-MFT}

\subsection{Hierarchical free energy}
\label{sec:Hierarchical-FE}

After performing $K$ scaling transformations (replications of spin
variables) we end up with $K$ geometric parameters $m_1<m_2<\ldots <
m_K=\nu$ as well as $K$ overlap susceptibilities
$\chi_1\ge\chi_2\ge\ldots\chi_K\ge0$ characterizing the phase space of the
order parameters. At each step (hierarchical level) of this construction we
use the Hubbard-Stratonovich transformation to linearize the newly
introduced (replicated) spin variables in free energy
\eqref{eq:FE-averaged-finite}. If we define $m_0=1$ and $\chi_{K+1}=0$ we
can represent the averaged free energy density with $K$ hierarchical
scaling transformations in the following form
  \begin{multline}\label{eq:avfe-density}
    f_K(q,\{\chi\};\{m\}) = -\frac\beta 4 (1-q)^2 + \frac \beta 4
    \sum_{l=1}^K (m_l-m_{l-1})\chi_l(2q + \chi_l) + \frac \beta2
    \chi_1\\ -\ \frac 1{\beta\nu} \int_{-\infty}^{\infty} \frac
    {d\eta}{\sqrt{2\pi}} e^{-\eta^2/2}
    \ln\left[\int_{-\infty}^{\infty}\frac {d\lambda_K}{\sqrt{2\pi}}
      e^{-\lambda_K^2/2}\left\{\dots \int_{-\infty}^{\infty}\frac
        {d\lambda_1}{\sqrt{2\pi}}
        e^{-\lambda_1^2/2}\right.\right. \\ \left. \left.
        \left\{2\cosh\left[\beta\left(h +
              \eta\sqrt{q} + \sum_{l=1}^{K}\lambda_l \sqrt{\chi_l -
                \chi_{l+1}}\right)\right]\right\}^{m_1}\ldots\right\}
      ^{m_K/m_{K-1}}\right]\ . \end{multline}
In this expression $q$ and $\chi_l, l=1,\ldots,K$ are physical order
parameters and are determined from the saddle-point equations. The numbers
$m_l,l=1,\ldots,K$ are formally external geometric parameters determining
the replica-symmetry breaking scheme of the matrix $\chi^{ab}$ from
Eq.~\eqref{eq:FE-averaged-finite}. They parameterize successive scalings
(replications) of extensive variables that would not change
thermodynamically homogeneous solutions.

It is evident that the averaged free energy density \eqref{eq:avfe-density}
can be uniquely analytically continued to arbitrary nonnegative numbers
$m_1,\ldots,m_K=\nu$, since it is represented by analytic functions for
all nonnegative variables $m_l,l=1,\ldots,K$. In the analytically
continued function, the geometric parameters $m_l$ need no longer be
integers and either they need not form an ascending sequence.

To find out whether a specific choice of geometric parameters $m_l$ can
lead to a thermodynamically homogeneous solution we have to understand how
the free energy depends on these parameters. We obtain from the structure
of the r.h.s. of Eq.~\eqref{eq:avfe-density} the following identities
\begin{subequations}\label{eq:degeneracy} \begin{align}
    f_K(q,\{\chi\};\{m, m_K=m_{K-1}\}) &= f_{K-1}(q,\{\chi\};\{m\}),
     \label{eq:m_degeneracy}
\\  f_K(q,\{\chi\};\{m, m_K=0\}) &=
f_{K-1}(q+\chi_K,\{\chi;\chi_i =\chi_i-\chi_K\};\{m\}),
\label{eq:0_degeneracy} \\  f_K(q,\{\chi, \chi_i=\chi_{i+1}\};\{m\}) &=
f_{K-1}(q,\{\chi\};\{m\}). \label{eq:chi_degeneracy}
\end{align}\end{subequations}
The solutions in the first and third cases are degenerate. The averaged
free energy density $f_K$ does not depend on $\chi_K$ in
Eq.~\eqref{eq:m_degeneracy} and it is independent of $m_i$ in
Eq.~\eqref{eq:chi_degeneracy}. Further on, if $m_{K} \ge m_{K-1}$ then
$f_{K} \le f_{K-1}$. We hence can conclude that if $f_K$ \emph{does
depend} on $m_K$, i.~e., it is inhomogeneous, the averaged free energy
displays a local \textit{maximum} for some value $m_K\in (0,m_{K-1})$.
Free energy $f_K(q,\{\chi\};\{m\})$ is independent of $\nu=m_K$ if and
only if $\chi_K=0$.

That is, we can find such a parameter $m_K$ at which the averaged free
energy density $f_K(q,\{\chi\};\{m\})$ reaches a local saddle point for
the given values of $m_1,\ldots,m_{K-1}$ with respect to variations of the
parameter $m_K$. Consequently, we can add the geometric parameters
characterizing the structure of the phase space to the variational
parameters. If the solution does not obey the global homogeneity condition
$\partial f_K(q,\{\chi\};\{m\})/\partial m_K \equiv 0$, the local
homogeneity condition $\partial f_K(q,\{\chi\};\{m\})/\partial m_K = 0$
then minimizes deviations from the global thermodynamic homogeneity. It
immediately follows from Eqs.~\eqref{eq:degeneracy} that this stationarity
point is a \emph{local maximum}. We fix in this way any new geometric
parameter emerging in the hierarchical construction and achieve a theory
with a thermodynamically determined ultrametric structure. Notice, that
both sets of parameters $\chi_1,\ldots,\chi_K$ and $m_1,\ldots,m_K$ form
sequences of decreasing numbers from interval $[0,1]$. It is easy to
verify that a substitution $q^{ab}=q+\chi^{ab}$ in
Eq.~\eqref{eq:avfe-density} recovers the Parisi RSB solution with $K$
discrete hierarchies.\cite{Dotsenko01}

\subsection{Hierarchical stationarity equations}
\label{sec:Stationarity-equations}

To simplify the analysis of properties of the hierarchical free energy
and the stationarity equations determining the variational parameters
$\chi_l$ and $m_l$ we rewrite the r.h.s. of Eq.~\eqref{eq:avfe-density} in
a recursive way. We define a sequence of partition functions \pagebreak[1]
\begin{equation}\label{eq:pf-hierarchy}
Z_l =  \left[\int_{-\infty}^{\infty}\mathcal{D}\lambda_l\
Z_{l-1}^{m_l}\right]^{1/m_l}
\end{equation}
with the initial condition $Z_0 = \cosh\left[\beta\left(h + \eta\sqrt{q} +
\sum_{l=1}^{K}\lambda_l \sqrt{\Delta\chi_l} + \lambda\ \sqrt{\Delta\chi}\
\right)\right]$. We denoted $\Delta\chi_l = \chi_l-\chi_{l+1}$ and
$\Delta\chi \equiv \chi_{K+1} = \chi - \sum_{l=1}^{K} \Delta\chi_l$. We
singled out the scaling parameter $\nu$ from the other geometric
parameters. The averaged free energy density can then alternatively be
represented as
 \begin{multline} \label{eq:mf-avfe}
f_K^\nu(q,\chi,\Delta\chi_1,\ldots,\Delta \chi_K; m_1,\ldots,m_K) =
-\frac\beta 4 (1-q -\chi)^2+ \frac \beta 4\nu\Delta\chi\left(2 q +
\Delta\chi\right)- \frac 1\beta \ln 2 \\ + \frac \beta 4 \sum_{l=1}^K
m_l\Delta\chi_l\left[2(q + \chi - \sum_{i=1}^{l-1}\Delta\chi_{i}) -
\Delta\chi_l\right] - \frac 1\beta \int_{-\infty}^{\infty}
\mathcal{D}\eta\
    \ln \left\{\int_{-\infty}^{\infty}\mathcal{D}\lambda\  Z_K^\nu \right\}^{1/\nu}
\end{multline}
with $q, \chi,\Delta\chi_l,l=1,\ldots,K$ and $m_l,l=1,\ldots,K$ as order
parameters to be determined from stationarity equations. The number of
hierarchies $K$ used in the free energy should be chosen so that
$f^\nu_K$ does not depend on the scaling parameter $\nu$.

To represent the mean-field equations we introduce a set of
hierarchical density matrices in the space of fluctuating random
fields $\lambda_l$. We define
$\rho_l(\eta,\lambda;\lambda_K,\ldots,\lambda_l) = Z_l^{m_l}/
\left\langle Z_l^{m_l}\right\rangle_{\lambda_l}$ and
$\rho(\eta,\lambda) = Z^{\nu}/\left\langle
  Z^{\nu}\right\rangle_{\lambda}$ with $Z= \left\langle
  Z_K^{m_K}\right\rangle_{\lambda_K}^{1/m_K}$.  We further introduce
short-hand notations $t \equiv \tanh\left[\beta\left(h + \eta\sqrt{q}
    + \lambda\sqrt{\Delta\chi} + \sum_{l=1}^K
    \lambda_l\sqrt{\Delta\chi_l} \right)\right]$ and $\langle
t\rangle_l(\eta,\lambda;\lambda_K,\ldots,\lambda_{l+1}) =
\langle\rho_l\ldots\langle\rho_1 t \rangle_{\lambda_1} \ldots
\rangle_{\lambda_l}$ with $\langle X(\lambda_l) \rangle_{\lambda_l} =
 \int_{-\infty}^{\infty}\mathcal{D}\lambda_l\
 X(\lambda_l)$.

With the above definitions we can write down the stationarity
equations for the physical order parameters
\begin{subequations}\label{eq:mfeqs-physical}
\begin{align}\label{eq:mfeqs-q}
  q(\nu,K) &= \langle\langle\rho\langle
  t\rangle_K\rangle_\lambda^2\rangle_\eta\\ \label{eq:mfeqs-chi}
  \chi(\nu,K) &= \langle\langle\rho\langle t^2
  \rangle_K\rangle_\lambda\rangle_\eta - \langle\langle\rho\langle t
  \rangle_K\rangle_\lambda^2\rangle_\eta\\ \label{eq:mfeqs:Delta_chi}
  \Delta\chi_l(\nu,K) &= \langle\langle\rho\langle \langle
  t\rangle_{l-1}^2 \rangle_K\rangle_\lambda\rangle_\eta -
  \langle\langle\rho\langle\langle t\rangle_{l}^2
  \rangle_K\rangle_\lambda\rangle_\eta
\end{align}
\end{subequations}
and for the geometric ones
\begin{align}\label{eq:mfeqs-geometric}
  m_l(\nu,K) &= \frac 4{\beta^2}\ \frac {\langle\langle\rho\langle \ln Z_{l-1}
    \rangle_K\rangle_\lambda\rangle_\eta - \langle\langle\rho\langle \ln
    Z_{l} \rangle_K\rangle_\lambda\rangle_\eta
    } {\langle\langle\rho\langle
    \langle t \rangle_{l-1}^2 \rangle_K\rangle_\lambda\rangle_\eta^2 -
    \langle\langle\rho\langle \langle t \rangle_{l}^2
    \rangle_K\rangle_\lambda\rangle_\eta^2}\
\end{align}
where index $l=1,\ldots,K$. A thermodynamically homogeneous solution is
obtained if $\chi= \sum_{l=1}^K\Delta\chi_l$ and the remaining
$2K+1$ order parameters do not depend on $\nu$.

\subsection{Stability conditions}
\label{sec:Stability}

Averaged free-energy density \eqref{eq:mf-avfe} defines a solution of the
SK model with $K$ hierarchical levels labeled by a scaling parameter
$\nu$. A globally thermodynamically homogeneous averaged free energy may
not depend on $\nu$. This happens if $\chi= \sum_{l=1}^K\Delta\chi_l$.
This condition of global thermodynamic homogeneity is satisfied
if an inequality
\begin{subequations}\label{eq:AT}
\begin{equation}\label{eq:AT-homogeneity}
 1 \ge \beta^2\left\langle \left\langle 1 -
   t^2 + \sum_{l=1}^{K} m_l \left(\langle
    t\rangle_{l-1}^2 - \langle t\rangle_l^2\right)\right\rangle_{K}^2
  \right\rangle_\eta
\end{equation}
is fulfilled. This inequality, however, does not represent the only
stability condition for a multilevel hierarchical free energy.  A
hierarchical solution with $K$ levels is stable if it does not decay into
a solution with $K+1$ hierarchies. A new order parameter
$\Delta\chi$ may emerge so that $\Delta\chi_l > \Delta\chi >
\Delta\chi_{l+1}$ for arbitrary $l$. That is, the new order parameter peels
off from $\Delta\chi_l$ and shifts the numeration of the order parameters
for $i>l$ in the existing $K$-level solution.  To guarantee that this does
not happen and that the averaged free energy depends on no more geometric
parameters than $m_1,\ldots,m_K$ we have to fulfill a set of $K$
generalized  AT stability criterions that for our hierarchical solution
read for $l=1,2,\ldots,K-1$
\begin{equation}\label{eq:AT-hierarchical}
 1 \ge \beta^2\left\langle\left\langle \left\langle 1 -
   t^2 + \sum_{i=1}^{l} m_i \left(\langle
    t\rangle_{i-1}^2 - \langle t\rangle_i^2\right)\right\rangle_{l}^2
  \right\rangle_K\right\rangle_\eta\ .
\end{equation}
There is also a condition that the new order parameter emerges as the
largest difference, that is $\Delta\chi > \Delta\chi_1$. So that neither
this instability takes place we have to fulfill
\begin{equation}\label{eq:AT-zero}
 1 \ge \beta^2\left\langle\left\langle \left( 1 -
   t^2 \right)^2  \right\rangle_K\right\rangle_\eta\ .
\end{equation}\end{subequations}
Actually, it is sufficient to take into account only a single stability
condition, namely that with the maximal right-hand side of
Eqs.~\eqref{eq:AT}. Which of these right-hand sides is maximal depends on
the particular choice of the optimal geometric parameters $m_1,\ldots,m_K$
minimizing thermodynamic inhomogeneity of the hierarchical solution with
lower numbers of hierarchical levels.

\subsection{Physical interpretation of the order parameters from the
hierarchical free energy} \label{sec:Hierarchical-interpretation}

The hierarchical free energy, Eq.~\eqref{eq:mf-avfe}, is equivalent to the
Parisi discrete RSB solution with $K$ hierarchies. Hence the resulting
stationarity equations, Eqs.~\eqref{eq:mfeqs-physical} and
\eqref{eq:mfeqs-geometric}, coincide with the stationarity equations
derived from the $K$-step RSB free energy when we make a substitution
$q^{ab} = q + \chi^{ab}$. The derived numbers must be the same. The
meaning and a physical interpretation of the order parameters in both
approaches may, however, be different. Different interpretations of the
role of the order parameters, in particular of the geometric ones,
originate from the  way the hierarchical free energy \eqref{eq:mf-avfe}
was derived. The RSB free energy was derived in an effort to maximize the
averaged free energy within the replica trick and the discrete scheme
\eqref{eq:mf-avfe} was used as an intermediate step toward its eventual
form -- the limit  $K\to\infty$ and continuously distributed order
parameters $q(x)$, $x\in[0,1]$. The physical interpretation of the Parisi
RSB solution is then based on this continuous limit.\cite{Mezard87} The
present approach does not provide justification for the continuous limit
and the physical meaning of the order parameters in the hierarchical
free energy must be sought within the discrete scheme.
 
The hierarchical free energy was derived by replica-symmetric averaging
of the TAP free energy extended to a replicated phase space. Real replicas
in the thermodynamic TAP approach were introduced to include  control
over thermodynamic homogeneity. Thermodynamic homogeneity is tightly
connected with uniqueness of equilibrium  states determined by mean-field
local magnetizations calculated from the TAP equations. The TAP equations
define a unique thermodynamic state if the solution reacts to all possible
perturbations identically and no possible internal structure of the
solution can be revealed. We showed that by replicating the phase space we
indeed reveal an internal structure of the TAP solutions.
 
It is clear from the construction itself that the overlap susceptibilities
$\chi^{ab}$ measure the interaction strength with which different copies
of spins thermodynamically influence each other. That is, the thermal
averaging of one spin copy depends on the values of spins in the other
copies if $\chi^{ab}>0$. We cannot separate individual replicas although
only one spin replica represents the physical system under consideration.
The non-replicated original phase variables together with temperature and
the chemical potential are hence insufficient to describe entirely the
equilibrium thermodynamic states. To get rid of the dependence of
thermodynamic states on boundary or initial  conditions we have to average
over all initial/boundary values and external variables that influence the
thermodynamics of the investigated system. In long-range, completely
connected models the degeneracy in solutions of the mean-field equations
is reflected in the dependence on the initial spin configurations. We
simulated this dependence in our approach  with replicas of the
spin variables subject to the same thermal equilibration.
 
If a mean-field solution is thermodynamically inhomogeneous, thermal
equilibration depends on the initial spin configuration. Dependence of the
thermodynamic state on the initial spin configuration from which we start
equilibration may have a nontrivial form. It is reflected in our approach
in the matrix $\chi^{ab}$. The replica-symmetric ansatz, $\chi^{ab} =
\chi$, means that all the initial spin configurations are equivalent and
that there is only a single "mean" strength with which they affect the
resulting equilibrium state. The replica symmetric ansatz is the most
natural first guess motivated by the high-temperature phase but need not
lead to a thermodynamically homogeneous solution. The dependence of the
equilibrium states on their initial configurations should be chosen so as
to reach a globally homogeneous solution. To achieve this goal we apply
successive replications using only the simplest, replica-symmetric ansatz
from the high-temperature solution at each stage. This construction seems
to be more transparent than an unjustified replica symmetry-breaking
ansatz. Moreover, the iterative construction offers an appealing physical
interpretation of the geometric order parameters used in the hierarchical
solution \eqref{eq:mf-avfe}.

To understand the role of the geometric parameters let us first take the
1RSB free energy~\eqref{eq:FE-1RSB}. The interacting part of the averaged
TAP free energy density reads
\begin{equation}\label{eq:TAP-change}
- \frac{N}{\beta V}\ln Z_0(\beta,h) \longrightarrow -\frac{1}{\beta\nu V}
\ln \left[\int\mathcal{D}\lambda\ Z_0\left(\beta, h + \lambda
\sqrt{\chi}\right)^\nu \right]^N\ .
\end{equation}
We can see that the replicated spins influence the original spins by making
the internal magnetic field dynamically random. The replicated free energy
then behaves as if  effectively $\nu N$ spins of the original system
enclosed in the volume $\nu V$ were affected by the replicated spins
outside the system under consideration. The internal magnetic field
changes due to the existence of replicated spins to $h\to h + \lambda
\sqrt{\chi}$. The integral over the fluctuating variable $\lambda$ stands
for averaging over the replicated (external) spins. The averaging over the
replicated spins is dynamical (annealed), in contrast to the quenched
(static) averaging over the random configurations of the spin exchange.
The parameter $\nu$ is then kind of a chemical potential governing the
exchange between the active and additional replicated spin configurations.
It has to be chosen so as the external replicated spins minimally influence
the final equilibrium state of our original system.

Adding more geometric order parameters with $1>m_1>\ldots>m_K>0$ in the
full hierarchical free energy means that the true equilibrium states are
hierarchically dependent on the initial spin configurations or they depend
on the history of quasi-equilibrium states they went through during thermal
averaging. When the thermodynamic equilibrium of the original system
depends on configurations of replicated spins we have to include the
replicated spins into our global thermodynamic system. In this merge we
compose the total number of $N$ spins from $m_1 N$ from the original
system and $(1 - m_1)N$ from the replicated one. The parameter $m_1$ is to
be chosen so as to minimize the impact of the replicated spins on the
original ones. Only the original spins, however, represent active
variables, while the replicated ones form kind of a thermal bath. We have
to iclude the bath explicitly into thermal equilibration of the active
spin variables, since the latter are affected by the former. The bath
spins affect the thermodynamics of the active spins via the overlap
susceptibility $\chi_1$ modifying their internal magnetic field. We
further replicate the $N$ spin variables in the whole volume and test
whether our $m_1 N$ active spins interacting with the bath with the
overlap susceptibility $\chi_1$ are affected by this replication. If yes,
we have to include the new replicated spins into thermal averaging. We
denote $\chi_2$ the strength with which the new (second level) replicas
affect the internal magnetic field of the active spins. The bath spins
(first level replicas) are then affected by the new spin replicas in the
same way as the first-level replicas act on the active spins, that is via
an overlap susceptibility $\chi_1 > \chi_2$. The optimal restructuring of
spins in the whole system is such that only $m_2 N$ spins belong to the
active ones, $m_1 N$ are from the first-level bath and the rest of $(1 -
m_1)N$ spins are the new replicated spins, second-level bath. The
parameters $m_1$ and $m_2$ are dynamically determined from minimization of
the impact of the newly replicated spins on the active ones. We continue
with wrapping the active spin variables in successively replicated ones so
long until we reach independence of the active spins on phase-space
replications. The hierarchical construction converges toward a globally
homogeneous solution if $\chi_1.> \chi_2>\ldots>\chi_K\to 0$, or
$\chi_{K+1} =0$  at a finite number of hierarchies~$K$.

Alternatively we can interpret the free energy
\eqref{eq:mf-avfe} as a solution with a multitude of equivalent
equilibrium thermodynamic states. Each state extends on average over a
portion $m_KN$ of the whole spin space. The states are organized
hierarchically with on average $(m_{K-1}-m_K)/m_K$ nearest neighbors with
the overlap susceptibility $\chi_1$, $(m_{K-2} - m_{K-1})/m_{K}$ next
nearest neighbors with the overlap susceptibility $\chi_{2}$ and so on.
The last $K$th level is characterized by $(1-m_1)/m_K$ neighbors with the
overlap susceptibility $\chi_K$. The total number of equilibrium states
then statistically is $1/m_K$. It is clear that  pure states cannot be
singled out and separated from their neighbors. Only the whole complex of
hierarchically arranged states can be thermodynamically homogeneous and
form an independent system with a well defined thermodynamic limit.

\section{Conclusions}
\label{sec:Conclusions}
 
We used the basic physical principle of thermodynamic homogeneity and
derived with the aid of real replicas in the thermodynamic TAP approach a
hierarchical representation for the averaged free energy of the SK model.
The hierarchical free energy \eqref{eq:mf-avfe} was derived via successive
replications of the phase space with the replica symmetric ansatz for the
introduced order parameters -- overlap susceptibilities. Real replicas
proved to be a suitable tool for treating situations when the TAP order
parameters, local magnetizations, do not describe unique thermodynamic
states. Real replicas enable one to lift the degeneracy of the TAP
approach and provide for a larger phase space within which a
thermodynamically homogeneous solution can be found. The replica symmetric
ansatz reflects an assumption of no internal structure (metric) of the
thermodynamic states corresponding to a given set of local magnetizations
from the TAP equations. This is the most natural (minimal) choice when we
do not know the actual structure and organization of equilibrium
states. Successive scalings in the phase space and the property of global
thermodynamic homogeneity then lead to a selection of a nontrivial,
ultrametric structure of thermodynamic states. We apply so many scalings
of extensive variables (hierarchical levels) until the global
thermodynamic homogeneity is achieved.
 
The averaged free energy \eqref{eq:mf-avfe} derived in this way is
equivalent to the Parisi discrete RSB solution. It contains a set of
averaged physical parameters $q,\chi_l$, $l=1,\ldots,K$ and a set of
geometric parameters $m_l$, $l=1,\ldots,K$. The geometric parameters are
turned variational ones  by the demand of local thermodynamic
homogeneity at each step of the hierarchical construction. The principle
of local thermodynamic homogeneity replaces the maximum principle in the
Parisi RSB construction. The homogeneity is reached successively by
demanding stability with respect to scalings of extensive variables.
Each hierarchical level then \textit{minimizes} deviations from the global
homogeneity and hence the instability of the solution. The maximum
principle of Parisi emerges as a consequence of minimization of
thermodynamic inhomogeneity of intermediate solutions and the form of
stationarity equations for the SK model. However, it does not mean that
the absolute maximum of the averaged free energy should be the equilibrium
solution. The maximum principle holds only for thermodynamically
inhomogeneous states. The free energy should still be minimal among
thermodynamically homogeneous states.

We were able to derive the discrete Parisi RSB scheme from a physical
principle of thermodynamic homogeneity but we do not find justification
for its continuous version characterized by a nonlinear differential
equation. The continuous version emerges in the limit $K\to\infty$ by
assuming infinitesimal smallness of the overlap susceptibilities,
$\chi_l=\Delta_l/K$ and infinitesimal differences in the geometric
parameters $m_l/m_{l+1}=1 + \delta_l/K$. In the continuous limit of the
RSB scheme the geometric parameters are no longer determined
thermodynamically, they cover interval $[0,1]$. Only an order-parameter
function $q(x)$ for $x\in [0,1]$ is to be determined variationally. In the
discrete scheme the geometric order parameters form a discrete set and are
determined thermodynamically from Eqs. \eqref{eq:mfeqs-geometric}. These
equations are essential part of the hierarchical solution and are of
particular importance at low temperatures. Only with thermodynamically
determined geometric parameters we are able to improve upon the SK free
energy at zero temperature. At low temperatures, new variational
parameters $x_l=\beta m_l$ are to be introduced and used instead of $m_l$.
Moreover, the thermodynamically shaped ultrametric structure of equilibrium
states in the discrete scheme leads to tangible nonlinear effects. They go
lost in the continuous limit. To decide whether the discrete or continuous
versions of the Parisi solution with $K=\infty$ holds in the SK model, one
has to evaluate the discrete scheme near the spin-glass transition point,
which has not yet been done. Work on comparison discrete and continuous
versions of the RSB near the critical temperature is in progress.

Thermodynamic inhomogeneity of the SK solution in the real-replica approach
was attributed to ambiguity of solutions of the TAP equations in the
determination of pure equilibrium states. Although real replicas offer a
way how to identify this degeneracy, they do not allow for separation of
individual pure states. That is why an organization of thermodynamic
states cannot be determined without an ansatz. We hence cannot find a
fully ansatz-free solution of the SK model. Successive scaling
transformations with the replica symmetric ansatz allow the system to
arrange equilibrium states so that thermodynamic inhomogeneity at
intermediate states is minimal. It recovers the Parisi discrete RSB
scheme, but we cannot claim that this is the only thermodynamically
homogeneous solution. At present, we cannot even prove that the full
(infinite level) solution is indeed thermodynamically homogeneous, that
is, it fulfills stability conditions \eqref{eq:AT}. Nevertheless, the
proposed construction seems to offer a rather straightforward way based on
basic principles of statistical mechanics to reach the discrete RSB
solution with stability conditions and an appealing physical
interpretation.

To conclude, we demonstrated that the discrete RSB solution of the SK model
is not a consequence of the replica trick and the limit of the number of
replicas to zero. Using real replicas we showed that the averaged free
energy is an analytic function of the number of replicas on the positive
axis. The thermodynamically formed ultrametric hierarchical structure with
$K$ levels of the order parameters in the SK model was shown to emerge due
to thermodynamic inhomogeneity of the replica symmetric solutions with
less than $K$ hierarchies. Thermodynamic homogeneity of the averaged free
energy with respect to scalings (replications) of the phase volume is
imposed at each hierarchical level. When not fulfilled, the free energy
depends on the geometric scaling factor that is then chosen in order to
minimize the inhomogeneity. It appears that in the SK model this
minimization leads to maximization of the free energy. The number of
hierarchical levels needed in this construction is fixed by the global
homogeneity condition, Eq.~\eqref{eq:AT-homogeneity}.

The work on this problem was supported in part by Grant IAA1010307 of the
Grant Agency of the Academy of Sciences of the Czech Republic and the ESF
Programme SPHINX. I thank Lenka Zdeborov\'a for fruitful discussions.


\begin{thebibliography}{99}
\bibitem{Sherrington75} D. Sherrington and S.  Kirkpatrick, \newblock
  Phys. Rev. Lett. {\bf 35}, 1972  (1975).

\bibitem{Parisi80} G. Parisi, \newblock J. Phys. A {\bf 13},
  L115, \textit{ibid} 1101, \textit{ibid} 1887, (1980).

\bibitem{Binder86} K.  Binder and A. P. Young, \newblock Rev. Mod.
  Phys. {\bf 58}, 801  (1986).

\bibitem {Mezard87} M. M\'ezard, G. Parisi, and M. A.  Virasoro,
  \newblock \textit{Spin Glass Theory and Beyond}, World Scientific
  (Singapore 1987).

\bibitem{Guerra02} F. Guerra,\newblock E-print \textit{cond-mat}/0205123.

\bibitem{Kirkpatrick78} S.  Kirkpatrick and D. Sherrington, \newblock
  Phys. Rev.  B{\bf 17}, 4384  (1978).

\bibitem{Talagrand04} M. Talagrand, \newblock preprint \textit{The Parisi
formula} to appear in Annals of Mathematics.

\bibitem{Parisi83} G.  Parisi, \newblock Phys.  Rev. Lett. {\bf 50},
   1946 (1983).

\bibitem{Janis87} V. Jani\v s, \newblock J. Phys. A
  {\bf 20},  L1017 (1987).

\bibitem{Franz93} S. Franz, G. Parisi, and M. A. Virasoro, \newblock
  J.  Phys.  I (Paris) {\bf 2} (1993) 1869.

\bibitem{Janis04a} V. Jani\v s and L. Zdeborov\'a, \newblock
\textit{cond-mat}/0407615, to appear in Progr. Theor. Phys.

\bibitem{Bray80} A. J. Bray and M. A. Moore, \newblock J. Phys. C:
Solid State Phys. {\bf 13}, L469 (1980).

\bibitem{Sommers78} H. J. Sommers, \newblock Z. Physik B{\bf 31}, 301
(1978).

\bibitem{Janis89} V. Jani\v s, \newblock Phys. Rev. B{\bf 40}, 11331
(1989).
 
\bibitem{Mezard86} M. M\'ezard, G. Parisi, and M. A. Virasoro,\newblock
Europhys. Lett. {\bf 1}, 77 (1986).

\bibitem{Plefka82} T. Plefka, \newblock J. Phys. A: Math. Gen. {\bf 15},
1971 (1982).

\bibitem{Plefka02} T. Plefka, \newblock Europhys. Lett. {\bf 58}, 892
(2002).
 
\bibitem{Janis90} V.  Jani\v s, \newblock Phys. stat. Solidi (b) {\bf
    157}, 425 (1990).


\bibitem{Monasson95} R. Monasson, \newblock Phys. Rev. Lett. {\bf 75}, 2847
(1995).

\bibitem{Mezard99} M. M\'ezard and G. Parisi, \newblock Phys. Rev. Lett.
{\bf 82}, 747 (1999).

\bibitem{Dotsenko01} V. Dotsenko, \textit{Introduction to the Replica
Theory of Disordered Statistical Systems}, Cambridge University Press,
(Cambridge 2001).

\end{thebibliography}
\end{document}